\documentclass[useAMS]{mn2e}
\usepackage{psfig}

\usepackage{times}
\usepackage{amssymb,amsmath}

\def\lsim{ \lower .75ex\hbox{$\sim$} \llap{\raise .27ex \hbox{$<$}} }
\def\gsim{ \lower .75ex \hbox{$\sim$} \llap{\raise .27ex \hbox{$>$}} }

\title[Extreme blazars and hadronic cascades] 
{On the hadronic cascade scenario for extreme BL Lacs}

\author[F. Tavecchio]
{Fabrizio Tavecchio\thanks{E--mail: fabrizio.tavecchio@brera.inaf.it}\\
INAF -- Osservatorio Astronomico di Brera, via E. Bianchi 46, I--23807 Merate, Italy\\
}

\begin{document}



\maketitle

\begin{abstract} 
The peculiar high-energy emission spectrum of the so-called extreme BL Lacs (EHBL) challenges the standard  emission models of blazars. Among the possible solutions, the so-called hadronic cascade scenario assumes that the observed high-energy radiation is produced in the intergalactic space through photo-hadronic reactions by ultra-high energy cosmic rays (UHECR) with energies up to $10^{19-20}$ eV beamed by the blazar jet. Under the assumption -- implicit in this model -- that the intrinsic high-energy synchrotron-self Compton emission of the blazar does not substantially contribute to the observed $\gamma$-ray spectrum, we derive constraints to the basic physical quantities of the jet and we compare them with the requirements of the hadronic cascade scenario. We found that, for a plausible range of  relativistic jet Doppler factors ($\delta=10-50$), the maximum achievable energy of the accelerated protons can exceed $2\times 10^{19}$ eV with jet powers of the order of $\approx 10^{44}$ erg s$^{-1}$, parameters compatible with the requests of the hadronic scenario even if EHBL are embedded in magnetic fields of cosmic filaments. We also discuss the consequences of our results for the possibility that  local EHBL contribute to the observed UHECR.
\end{abstract}
 

\section{Introduction}

Blazars constitute by far the dominant population of extragalactic $\gamma$-ray sources, both at GeV 
(e.g., Ackermann et al. 2011) and TeV (e.g., Rieger et al. 2013) energies. Their powerful, variable non-thermal emission, extending from the radio to the $\gamma$-ray band, originates in a relativistic jet pointing toward the observer. The broad band spectral energy distribution (SED) reveals the presence of two broad ``humps". The peak at low energies (with maximum at IR-X-ray frequencies, depending on the source) is due to synchrotron radiation by relativistic electrons.  The second component (peaking at MeV-GeV energies) is widely attributed to inverse Compton (IC) emission by the same electron population. For BL Lac objects, the majority of  blazars detected at TeV energies, characterized by the extreme weakness of emission lines in the optical spectra, is popularly assumed that the target soft photons for the IC scattering are the synchrotron photons themselves (synchrotron self-Compton, SSC, e.g. Tavecchio et al. 1998, TMG98 hereafter). Alternatively, hadronic scenarios model the high-energy component as the left-over of electromagnetic cascades initiated by high-energy protons (Mannheim 1993, Muecke et al. 2003) or as their direct synchrotron emission (Aharonian 2000). 
The necessity to accelerate protons at very high energies, together with the relatively low efficiency of the photo-meson and synchrotron proton emission (e.g., Sikora 2011), generally implies magnetic fields and jet powers larger than those inferred in leptonic models (B{\"o}ttcher et al. 2013). Moreover, the rapid variability observed in $\gamma$ rays, down to few hours in GeV (e.g. Saito et al. 2013, Foschini et al. 2013) and few minutes in TeV (e.g. Aharonian et al. 2007a, Albert et al. 2007, Aleksic et al. 2011), seems difficult to be reconciled with the inefficient cooling provided by hadronic interactions. 

A possible hadronic origin of the $\gamma$--ray component  has been recently considered for a small peculiar group of TeV emitting blazars, the so-called extreme high-energy peaked BL Lacs (EHBL, Costamante et al. 2001). The TeV spectrum of these sources (up to $\sim 10$ TeV), if corrected for the expected substantial depletion of high-energy photons caused by the interaction with the IR-optical extragalactic background light (EBL), is exceptionally hard (Aharonian et al. 2007b). Furthermore, in evident contrast with the rest of the blazar population, the VHE emission seems to be stable on timescales of $\sim$months-years, although there is room for small-amplitude variability, very difficult to detect because of the very low flux (e.g. Cerruti et al. 2013). These features make very difficult to interpret the SED in the framework of the standard SSC model, in particular because the reduction of the  Klein-Nishina scattering cross-section necessarily entails soft spectra above 1 TeV. 
Proposed solutions to this problem involve very high values of the minimum Lorentz factor of the emitting electron (Katarzy{\'n}ski et al. 2006, Tavecchio et al. 2009), Maxwellian-like electron energy distribution (Lefa et al. 2011), internal absorption in the source (Aharonian et al. 2008), IC emission in the Thomson regime in the large-scale jet (B{\"o}ttcher et al. 2008). In yet another, recent, possibility one speculates that high-energy photons do not originate in the blazar jet but, instead, are produced in the intergalactic space by ultra-high energy cosmic rays (UHECR) accelerated in the jet and beamed toward the observer. These traveling UHECR  loose energy through photo-meson and pair creation (Bethe-Heitler) reactions, triggering the formation of electromagnetic cascades in the intergalactic space. The $\gamma$ rays released in the cascades suffer less and less absorption as the cascade proceeds and the distance to the Earth decreases (e.g., Essey \& Kusenko 2010, Essey et al. 2011, Takami et al. 2013).

Razzaque et al. (2012) and Murase et al. (2012) recently explored in detail the hadronic cascade scenario for EHBL,  deriving the UHECR power required to account for the observed high-energy component, taking into account the possible deflection of  UHECR by the extragalactic (EGMF) and the intergalactic (IGMF) magnetic fields. In these studies the acceleration of UHECR in jets of EHBL is assumed as a starting point. Based on the physical parameters inferred for nearby ``normal" BL Lacs, Murase et al. (2012) assumed that the maximum energy of UHE protons can barely reach $E\sim 10^{19}$ eV. Protons up to this energy are inevitably substantially deflected by EGMF, leading to request large power in UHECR. This conclusion motivates our work, aimed to explore if the peculiar emission of EHBL implies physical conditions more favorable to satisfy the requirement of the hadronic model, in terms of maximum UHECR energy and power. Indeed, given the large ratio between the synchrotron and the high-energy peak luminosities, one could expects relatively large magnetic fields, and thus high UHECR energies. Large enough UHECR energies would limit the deflection by magnetic fields and therefore the required power of the beam to values suitable for EHBL jets. 

In \S 2 we derive constraints to the jet parameters and to the maximum energy of accelerated UHECR through a general analytical treatment  of the synchrotron-SSC one-zone model. In \S 3 we perform a detailed SED modeling of EHBL SED. In \S 4  we discuss the results considering also the possibility that a population of nearby EHBL could account for the observed local UHECR. 

We use $H_{\rm 0}\rm =70\; km\; s^{-1}\; Mpc^{-1} $, $\Omega_{\Lambda}=0.7$, $\Omega_{\rm M} = 0.3$.

\begin{figure*}
\hskip 1 cm
\psfig{file=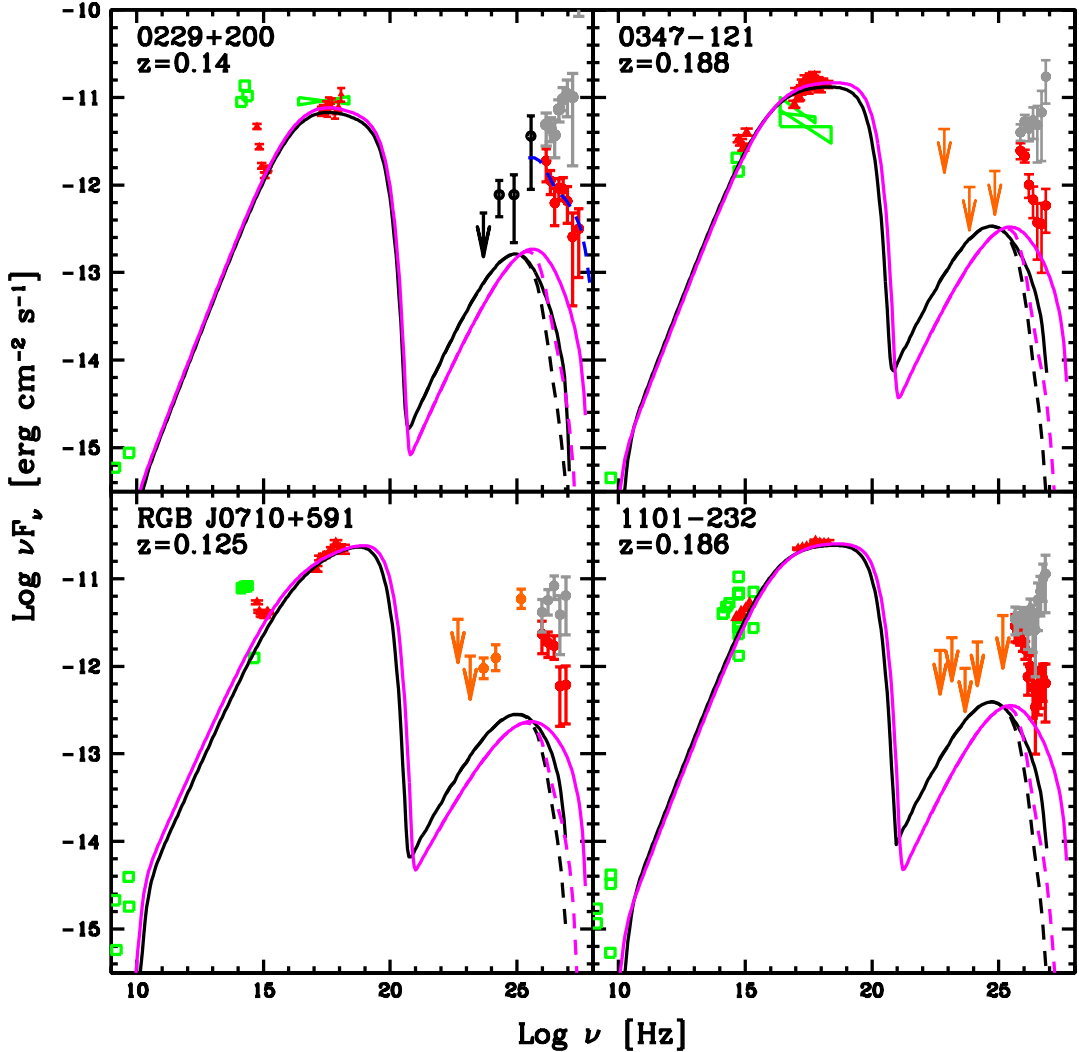,height=15cm}
\caption{SED of the EHBL analyzed in this work. Red symbols show {\it Swift} (XRT and UVOT) data, orange points report {\it Fermi}/LAT data (detections or upper limits from 2FGL catalogue, Nolan et al. 2012), red (gray) filled circles show the observed (de-absorbed) TeV data. Black circles for 1ES 0229+200 show LAT data from Vovk et al. (2012). Green symbols show historical data. For 1ES 1101-232 we used UV-optical data from Costamante (2007). See Tavecchio et al. (2010) for  references. The dashed blue line in the 1ES 0229+200 panel shows the theoretical spectrum from the hadronic cascade derived by Murase et al. (2012). The solid curves show the result of the emission model for Doppler factor of $\delta=10$ (black) and $\delta=30$ (magenta). The dashed curves at TeV energies is the spectrum including absorption through interaction with the EBL (using the Dominguez et al. 2011 model).}
\label{sed}
\end{figure*}

\section{General analytical constraints}

Fig. \ref{sed} reports the SED of the four  EHBL with the best sampling of the SED: 1ES 0229+200, 1ES 0347-121, RGB J0710+591 and 1ES 1101-232 (details on the data in the figure caption). Remarkably, the shape of the SED and the luminosities are very similar. In particular, the SED displays an extreme X-ray to radio flux ratio, which can be used as an effective discriminator in the EHBL selection (e.g., Bonnoli et al., in prep). In the optical band the emission is dominated by the thermal radiation of the host galaxy, while the jet non-thermal continuum starts to dominate only at UV wavelengths. The X-ray spectrum is hard, locating the synchrotron peak in or beyond the X-ray band.

The flux in the GeV band is rather low, hence difficult to be revealed by {\it Fermi}/LAT. Indeed only 1ES 0229+200 (Vovk et al. 2012) and RGB J0710+591 (Acciari et al. 2010) have been detected; for the other two sources only upper limits have been derived. At higher energies the observed TeV spectrum (red symbols) is soft: however, if the correction for absorption through the interaction with EBL is applied, very hard spectra (gray points) are derived. 






For the simple one-zone leptonic model, robust general constraints to the physical parameters of the blazar emission region can be derived (TMG98). In turn, the inferred parameters can be used to calculate the relevant quantities for the UHECR scenario, the maximum energy of the accelerated particles (using the so-called Hillas criterion) and the jet power. In this Section the constraints are derived and compared with the requests of the hadronic cascade scenario. 

We remark that the characteristics of the emission of EHBL are rather peculiar among the TeV emitting blazars. Indeed, ``classical" BL Lac detected at TeV energies show softer X-ray spectra, locating the synchrotron peak in the UV-soft X-ray band. Moreover, the GeV emission of classical BL Lacs is much more powerful than that of EHBL, implying a lower synchrotron/IC luminosity ratio. For these reasons, the application of the parameters  estimated for jets of classical BL Lacs to EHBL could be not appropriate.

The assumption at the base of  our procedure is that the intrinsic (EBL-absorbed) SSC emission of the jet does not substantially contribute to the observed high-energy component (which is instead the result of the hadronic cascade). The constraints derived under this assumption will not be limited to a specific (leptonic) emission framework. Indeed, the relativistic electrons emitting the synchrotron radiation will inevitably emit SSC radiation, making our constraints meaningful even if there are other (i.e. IC emission with external photon fields, hadronic) contributions to the $\gamma$-ray emission. Since these other contribution would imply larger magnetic fields and jet power, the values derived here could be considered as robust {\it lower limits} for these quantities. It is also worth noting that even if the UHECR acceleration occurs in regions different from that in which the observed synchrotron radiation is produced, the constraints to the jet power (though not those for $E_{\rm max}$) remains valid as long as the jet is assumed stationary.\\

\subsection{Maximum energy} 

The SSC to synchrotron luminosity ratio, $\mathcal{L}_C/\mathcal{L}_s$ is equal to the ratio of the  (comoving) synchrotron radiation and magnetic field energy density. 
In the Thomson limit for the IC scattering we can write :
\begin{equation}
\frac{\mathcal{L}_C}{\mathcal{L}_s}=\frac{U^{\prime}_{s}}{U^{\prime}_B} \; \rightarrow \; \frac{\mathcal{L}_C}{\mathcal{L}_s}
\simeq \frac{2 f \, \nu_s\mathcal{L}(\nu_s)}{B^2 R^2 c \, \delta^4}
\label{lumrat}
\end{equation}
\noindent
in which the energy densities are measured in the comoving frame of the source, $B$, $R$ and $\delta$ are the (comoving) magnetic field intensity and the radius of the (spherical) emission region and the Doppler factor\footnote{$\delta=[\Gamma(1-\beta \cos\theta_{\rm v})]^{-1}$, where $\Gamma$ is the bulk Lorentz factor, $\beta=v/c$ and $\theta _{\rm v}$ is the viewing angle. For $\theta_{\rm v}=1/\Gamma$, $\delta=\Gamma$ as assumed in this work.}. The synchrotron bump is generally well described by two power laws with slopes $\alpha_{1,2}$. The numerical factor $f$,  function of the spectral indices (see Appendix), connects the total  (i.e. frequency integrated) luminosity $\mathcal{L}_s$ to the SED peak luminosity $\nu_s\mathcal{L}(\nu_s)$ (TMG98). For simplicity in the following we use the notation $L\equiv\nu \mathcal{L}({\nu})$ to indicate the synchrotron and IC peak luminosities.

Following Hillas (1984), the maximum energy  (measured in the observer frame) of ions with charge $Ze$, accelerated in a source with magnetic field $B$, radius $R$ and bulk Lorentz factor $\Gamma$, is $E_{\rm max}\simeq  Z e B R \, \Gamma$. Combined with Eq. \ref{lumrat} and inserting the numerical values we obtain:
\begin{equation}
E_{\rm max}=2.6\times10^{19} \, \frac{Z}{\delta_1} \,\frac{L_{s,44}}{L_{C,42.5}^{1/2}}\frac{\Gamma}{\delta} \; \; {\rm eV},
\end{equation}
\noindent
where we introduced the notation $Q_X\equiv Q/10^X$. We calculate $f=11.4$ with $\alpha_1=0.3$ and $\alpha _2=1.1$, suitable to model the synchrotron peak of EHBL. The synchrotron luminosity is specialized to the value observed in EHBL, while $L_C$ is normalized to a value ten times smaller than the value of the luminosity of the high energy component,  tracked by the GeV and TeV data, $L_{\gamma}\sim 3\times 10^{43}$ erg s$^{-1}$.

Interestingly, $E_{\rm max}$  depends only on the synchrotron and SSC output and on the Doppler factor of the jet. Since $L_s$ is observed and $L_C$ can be limited, $E_{\rm max}$ depends only on the Doppler factor. The inverse proportionality between $E_{\rm max}$ and $L_C$ is easily understood considering that, for a fixed synchrotron luminosity, a lower $L_C$ implies a larger magnetic field (and viceversa) and thus a larger $E_{\rm max}$.

Given the high peak frequencies typical of EHBL, the SSC component is likely produced in the Klein-Nishina regime. In this case the correct expression includes also a (weak) dependence on the synchrotron and SSC peak frequencies, $\nu_s$ and $\nu_C$. Following TMG98 (see Appendix) we obtain (here and in the following we neglect the redshift correction given the small $z$ of the considered sources):
\begin{equation}
E_{\rm max}=3\times10^{19} \, \frac{Z}{\delta_1^{\alpha_1}} \, (\nu_{s,17} \nu_{C,25})^{(\alpha_1-1)/2}\,\frac{L_{s,44}}{L_{C,42.5}^{1/2}}\frac{\Gamma}{\delta} \; \; {\rm eV}
\label{emaxkn}
\end{equation}
\noindent
where we used again $\alpha_1=0.3$ and $\alpha _2=1.1$.

\begin{figure}
\hskip -0.6 cm
\psfig{file=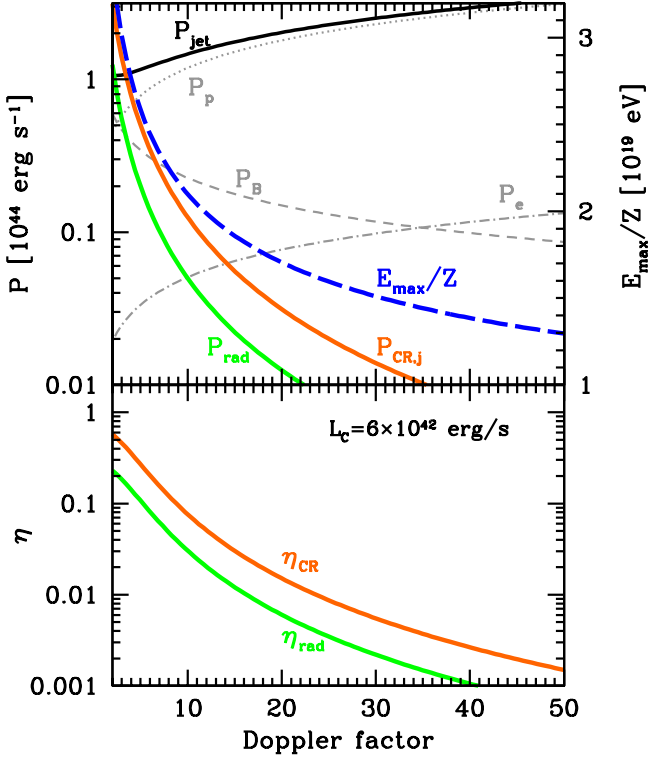,height=10.1cm}
\caption{{\it Upper panel:} the total jet power (black solid line) versus the jet Doppler factor. The gray lines separately show the contribution of the magnetic field (dashed), relativistic electrons (dot-dashed) and cold protons (dotted). The latter two contributions are calculated assuming an emission region with radius $R=10^{16}$ cm and a lower limit for the electron energy distribution $\gamma_{\rm min}=1$. The orange solid line shows the beaming-corrected power channeled into UHECR assuming an isotropic equivalent luminosity of $L_{\rm CR}=5\times 10^{45}$ erg s$^{-1}$. The green solid line shows the beaming-corrected radiative power of the jet. The blue dashed line reports the maximum energy of UHECR accelerated in the jet using the Hillas condition (scale on the left). {\it Lower panel}: efficiency in the production of UHECR (orange) and radiation (green) as a function of the jet Doppler factor.}
\label{emaxdelta}
\end{figure}

\begin{figure}
\hskip -0.6 cm
\psfig{file=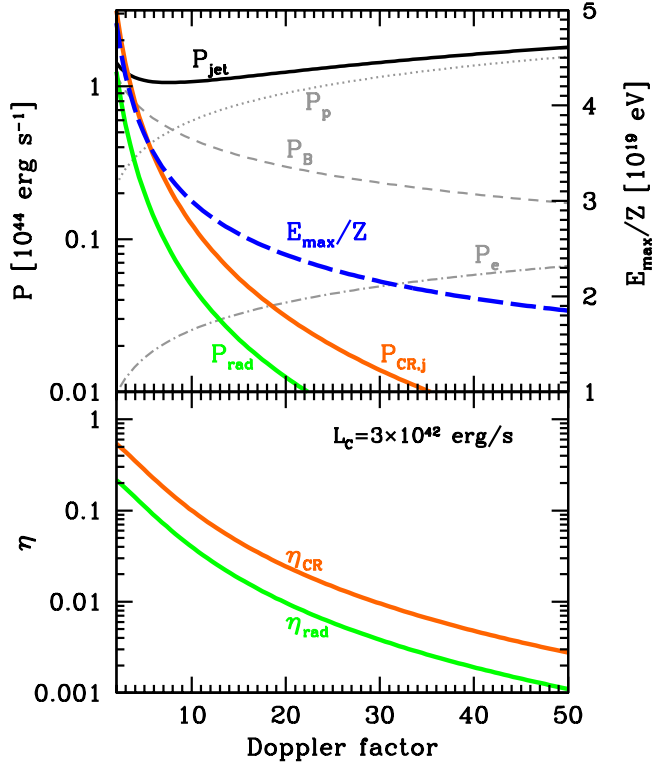,height=10.1cm}
\caption{As in Fig.\ref{emaxdelta} but for a SSC luminosity $L_C=3\times 10^{42}$ erg s$^{-1}$. Note the different scale for $E_{\rm max}$. 
}
\label{emaxdelta2}
\end{figure}

Acceleration of protons up to $\sim 2-3\times 10^{19}$ eV or iron ($Z=26$) up to $\sim 5\times 10^{20}$ eV is possible for standard ($\delta>10$) Doppler factors (we assume $\Gamma=\delta$).  In principle, even larger values would be possible allowing for a larger ratio between synchrotron and SSC luminosities. In Figs. \ref{emaxdelta}-\ref{emaxdelta2} we show (blue dashed lines) $E_{\rm max}/Z$ as a function of $\delta$ for two values of the SSC luminosity, $L_C=6\times 10^{42}$ and $3\times 10^{42}$ erg s$^{-1}$. We remark that the weak dependence on $\delta$ makes the estimate of $E_{\rm max}$ rather insensitive on this parameter.\\

The Hillas criterion, stating the equality of the gyration radius of the accelerating UHECR and the source size, should be considered an upper limit. Depending on the details of the acceleration mechanism and on the losses suffered during acceleration, a lower maximum energy could be achieved in real cases.



The acceleration timescale (in the source frame) is generally expressed in terms of the gyration time $t^{\prime}_{\rm g}=2\pi r_{\rm L}/c=2\pi E_{\rm p}^{\prime}/ZeBc$, including a factor $\epsilon >1$ incorporating the details of the mechanism:
\begin{equation}
t^{\prime}_{\rm acc}(E^{\prime}_{\rm p})=\epsilon \, \frac{2\pi E^{\prime}_{\rm p}}{ZeBc}\simeq 7\times 10^5 \epsilon \, E^{\prime}_{p,18} B^{-1}\,\, {\rm s}.
\label{acc}
\end{equation}
As shown in \S 2.3, radiative and photo-pion losses are negligible for protons in EHBL and the only relevant losses are due to the adiabatic expansion, for which the timescale is (e.g. Sikora 2011): 
\begin{equation}
t^{\prime}_{\rm ad}=\frac{3}{2} \frac{R}{\theta_{\rm j}\Gamma c}\simeq 5\times 10^5 \frac{R_{\rm 16}}{\theta_{\rm j}\Gamma} \,\, {\rm s},
\label{cool}
\end{equation}
where $\theta_{\rm j}$ is the opening angle of the jet. Recent VLBI observations of blazar jets (Clausen-Brown et al. 2013, Jorstad et al. 2005, Pushkarev et al. 2009) suggest $\theta_{\rm j}\Gamma\approx 0.2$.

Equating acceleration and cooling timescale with the values above and boosting it in the observer frame we find the maximum energy:
\begin{equation}
E_{\rm max,acc}=\frac{3.5\times 10^{19}}{\epsilon} Z R_{16} B \, \Gamma _1 \,\, {\rm eV}.
\label{emaxacc}
\end{equation}

The value of $\epsilon$ is highly uncertain. For diffusive shock acceleration it depends on the shock structure and on the details describing the magnetic turbulence (e.g. Rieger et al. 2007, Bykov et al. 2012). If acceleration is powered by magnetic reconnection, $\epsilon \approx 2$ is possible for reasonable reconnection rates (Giannios 2010). In this case the maximum energy roughly coincides with the ideal Hillas limit.

\subsection{Jet power and UHECR luminosity} 

The other key fundamental quantity for the UHECR initiated cascade scenario is the power of the jet. Murase et al. (2012) found that relatively large values of the isotropic equivalent luminosity in UHECR, $L_{\rm CR}$, are needed to reproduce the observed TeV spectrum of 1ES 0229+200. The necessary CR luminosity critically depends on the possible deviation of the UHECR by magnetic fields, both in the filaments/clusters in which the blazars resides (EGMF), and in the voids crossed by the traveling CR (IGMF). According to Murase et al. (2012), the deviation is negligible in the filaments case for protons of energy $E\gtrsim 2\times 10^{19}$ eV, while for clusters, only protons with energy exceeding $\sim 10^{21}$ eV suffer small deviation. Heavier ions are deflected by angles $Z$ times larger than protons of the same energy.

For the case of 1ES 0229+200, Murase et al. (2012) found that the minimum power requirement (negligible deviation by EGMF and IGMF) corresponds to an isotropic equivalent (i.e. not beaming corrected) luminosity of $L_{\rm CR}\sim 10^{45}-10^{46}$ erg s$^{-1}$ (depending on the details of the spectrum), which translates into an absolute, beaming corrected, power of $P_{\rm CR,j}=L_{\rm CR}/4\, \Gamma^2$ (where we assume a beaming angle $\theta_{\rm b}=1/\Gamma$ and we consider only one jet). Given the similar distance and $\gamma$-ray luminosity of all the EHBL considered here, we assume that this estimate is valid for all the sources. Obviously, in order to the hadronic model to be applicable, the power carried by the jet should be enough to sustain the required UHECR luminosity. For definiteness, in the following we fix $L_{\rm CR}=5\times 10^{45}$ erg s$^{-1}$.

In general, the jet power is the sum of the magnetic, relativistic electrons and, if present,  proton components (e.g. Celotti \& Ghisellini 2008). The magnetic (or Poynting) power can be expressed as $P_B=\pi R^ 2 \beta c \,U^{\prime}_B \Gamma^2$ that, using Eq. A1, can be written as: 
\begin{equation}
P_B=\frac{2\times 10^{44}}{\delta^{2\alpha_1}} \,(\nu_{s,17}\nu_{C,25})^{\alpha_1-1}\,\frac{L_{s,44}^2}{L_{C,42.5}} \left(\frac{\Gamma}{\delta}\right)^2 {\rm erg \, s^{-1}}
\label{power}
\end{equation}
\noindent
Like $E_{\rm max}$, also this quantity depends only on the SED parameters and $\delta$.

The kinetic power carried by the jet in the relativistic electron component, $P_{\rm e}$ is
$P_{\rm e}=\pi R^ 2  U_{\rm e}^{\prime} \beta c \, \Gamma^2$
where $U_{\rm e}^{\prime}$ is the electron energy density (including the rest mass energy): 
\begin{equation}
U_{\rm e}^{\prime}=m_e c^2 \int n(\gamma) \, \gamma \, d\gamma
\label{ue}
\end{equation}
Assuming a power-law  electron energy distribution, $n(\gamma)=k\gamma^{-p}$ with $\gamma_1<\gamma<\gamma_2$, $\gamma_2>>\gamma_1$ and $p<2$ (suitable for EHBL, see below) we have:
\begin{equation}
U_{\rm e}^{\prime}=\frac{m_e c^2k}{2-p}\, \gamma_2^{2-p}. 
\label{ue2}
\end{equation}
The electron density normalization $k$ can be derived from the total (i.e. frequency integrated) synchrotron luminosity $\mathcal{L}_s$:
\begin{equation}
\mathcal{L}_s\simeq\frac{4}{3} \sigma_{\rm T} c \frac{B^2}{8\pi} \mathcal{V} \delta^4 \int n(\gamma)\gamma^2 d\gamma
\label{ls}
\end{equation}
\noindent
where $\sigma _{\rm T}$ is the Thomson cross-section and $\mathcal{V}=(4/3)\pi R^3$ is the (assumed spherical) emitting volume. Assuming, as above, a power law electron energy distribution, the integral can be calculated as $k\gamma_2^{3-p}/(3-p)$. Using Eq. \ref{ls} and \ref{ue2} and recalling the definition of the magnetic power $P_B$, $P_{\rm e}$ can be expressed as:
\begin{equation}
P_{\rm e}=\frac{9\pi m_e c^3}{16\sigma_{\rm T}} R \, \frac{\mathcal{L}_s}{P_B}\, \frac{3-p}{2-p} \frac{1}{\gamma_2} \left( \frac{\Gamma}{\delta}\right)^4
\end{equation}

Analogously, the contribution of cold protons, with numerical density assumed equal to that of the electrons, can be calculated as:
\begin{equation}
P_{\rm p}=\frac{9\pi m_p c^3}{16\sigma_{\rm T}} R \, \frac{\mathcal{L}_s}{P_B}\, \frac{3-p}{1-p} \frac{\gamma_1^{1-p}}{\gamma_2^{3-p}} \left( \frac{\Gamma}{\delta}\right)^4.
\end{equation}
Note that both $P_e$ and $P_p$  are directly proportional to the SSC
luminosity and inversely proportional to the synchrotron luminosity (through their dependence on $P_B$) and  they also depends on the radius of the emitting region.

The total jet power $P_{\rm jet}=P_B+P_e+P_p$ is reported in Figs. \ref{emaxdelta}-\ref{emaxdelta2} (black solid lines), together with the separate components (gray lines). We present two cases, $L_C=6\times 10^{42}$ erg s$^{-1}$ and $L_C=3\times 10^{42}$ erg s$^{-1}$. We assumed values of the parameters suitable to describe EHBL (see next section), namely $p=1.6$, $\gamma_2=2\times 10^4$. Recalling (see Eq. 1) that $\mathcal{L}_s = f \times \nu _s \mathcal{L}(\nu _s)$, we use $\mathcal{L}_s =10^{45}$ erg s$^{-1}$.
Moreover we fix $\gamma_1=1$, although this quantity is poorly constrained by the SED (see below). Larger values would imply lower proton ($P_{\rm p}\propto \gamma_{1}^{-0.6}$) and (slightly) lower electron powers. We  assume a ``typical" radius $R=10^{16}$ cm (e.g., Tavecchio et al. 2010) for $P_e$ and $P_p$. The total jet power, but for the lowest values of $\delta$, is dominated by the proton component.

We also show in Fig. \ref{emaxdelta}-\ref{emaxdelta2} the beaming-corrected radiative luminosity of the jet, $P_{\rm rad}=(\mathcal{L}_s+\mathcal{L}_C)/2\Gamma^2\sim \mathcal{L}_s/2\Gamma^2$,  and the absolute CR luminosity in case of negligible CR deviation, $P_{\rm CR,j}=L_{\rm CR}/4\, \Gamma^2$ (in both cases we assume $\Gamma=\delta$). In the lower panel we report the corresponding efficiencies, $\eta_{\rm rad}=P_{\rm rad}/P_{\rm jet}$ and $\eta_{\rm CR}=P_{\rm CR,j}/P_{\rm jet}$. Note the rather similar radiative and CR efficiency. Low efficiencies in the CR acceleration ($\eta_{\rm CR}<0.1$) are required for $\delta>10$ to satisfy the required UHECR energetics.

\begin{table}
\begin{center}  
\begin{tabular}{lccccccc} 
\hline  
      &$ \gamma_{\rm b}$ & $\gamma _{\rm max}$& $n_1$&  $n_2$ & $B$ & $K$ &  $\delta $ \\
      & [$\times10^4$]& [$\times10^6$]& &  & [G]& [cm$^{-3}$] &\\
\hline 
0229+200 & $5.2$ & $2$& 1.5& 3.2& 0.1& 18.5& 30 \\
                         & $3.2$ & $1.2$& 1.5& 3.2& 0.75& 4.9& 10  \\
0347--121 & $2.8$ & $2$& 1.5& 3.0& 0.2& 3.2& 30  \\
                         & $1.5$ & $1$& 1.5& 3.0& 1.5& 10.0& 10 \\
1101-232 & $2.8$ & $2$& 1.5& 3.0& 0.29& 2.4& 30 \\
                         & $1.5$ & $1$& 1.5& 3.0& 2.3& 7.6& 10 \\
0710+591 & $2.5$ & $1.8$& 1.7& 2.85& 0.19& 13.0& 30 \\
                         & $2.5$ & $1$& 1.7& 2.85& 1.13& 32.0& 10 \\
\hline \\
\end{tabular}
\end{center}
\caption{Input parameters for the emission models shown in Fig.\ref{sed}. See text for definitions. In all the cases a radius $R=10^{16}$ cm and a minimum Lorentz factor of the electrons $\gamma_{\rm min}=1$ have been used. 
}
\label{model}
\end{table}

\begin{table*}
\begin{center}  
\begin{tabular}{lcccccccc} 
\hline  
      & $\Gamma $ & $\log P_p$ &$\log P_e$ & $\log P_B$ & $\log P_r$ & $\eta_{\rm rad}$ & $\eta_{\rm CR}$  & $E_{\rm max}/Z$\\
      & & [erg s$^{-1}$] & [erg s$^{-1}$] & [erg s$^{-1}$] & [erg s$^{-1}$] & & & [eV]\\
\hline 
0229+200  & 30 & 43.7 & 43.1 & 42.5 & 41.6 &  $6\times 10^{-3}$    & 0.02    & $9.0\times 10^{18}$\\
                  & 10& 43.1 & 42.5 & 43.3 & 42.6&    0.11   &  0.35   & $2.2\times 10^{19}$\\
0347--121 & 30& 43.9 & 43.2 & 43.1 & 42.2&   0.015   & 0.013    & $1.8\times 10^{19}$\\
                  & 10& 43.4 & 42.6 & 43.9 & 43.1&   0.12   &  0.12   &$4.5\times 10^{19}$\\
1101-232  & 30& 43.8 & 43.1 & 43.5 & 43.0&   0.09   &   0.01  & $2.6\times 10^{19}$\\
                 & 10& 43.3 & 42.5 & 44.3 & 44.0 &  0.45    & 0.06    &$6.9\times 10^{19}$\\
0710+591 & 30& 44.4 & 43.0 & 43.1 & 42.6&   0.01   & $5\times 10^{-3}$    & $1.7\times 10^{19}$\\
                 & 10& 43.8 & 42.4 & 43.7 & 43. 6&    0.34  &  0.11   &$3.4\times 10^{19}$\\
\hline \\
\end{tabular}
\end{center}
\caption{Quantities inferred from the spectral models. We report the power carried by the jet as (cold) protons, electrons, magnetic field, the efficiency in radiation and UHECR and $E_{\rm max}$. $\Gamma=\delta$ is assumed. The UHECR efficiency is estimated assuming an isotropic equivalent luminosity $L_{\rm CR}=5\times 10^{45}$ erg s$^{-1}$.}
\label{model}
\end{table*}

As a general trend, larger $E_{\rm max}$ are obtained for lower Doppler factors. On the contrary, larger values of $\delta$, determining a smaller beaming cone, imply lower CR powers and lower $\eta_{\rm CR}$.

Note finally that $P_B>P_{\rm CR,j}$, implying that the magnetic pressure always dominates over the pressure of the UHECR. This condition guarantees that UHECR can be effectively confined by magnetic field during the acceleration.

\subsection{UHECR synchrotron losses} 

So far we have not considered that UHECR in the jet loose energy through synchrotron emission before to leave the source\footnote{As already noted by Murase et al. (2012), photo-pion losses are completely negligible for protons in jets of BL Lacs due to the low photon target energy density.}. Synchrotron cooling can have a twofold effect: 1) if synchrotron losses are important, they could limit the maximum achievable energy $E_{\rm max}$; 2) the synchrotron emission from UHECR, peaked in the $\gamma$-ray band, could be not negligible and thus contribute to the total emission. 

The (comoving) cooling time for relativistic protons of (comoving) energy $E^{\prime}_{\rm p}$ is:
\begin{equation}
t^{\prime }_{\rm p}(E^{\prime}_{\rm p})=\left( \frac{m_{\rm p}}{m_{\rm e}} \right)^2 \frac{6\pi \, m_{\rm p}^2 c^3}{\sigma_{\rm T} B^2 \, E^{\prime}_{\rm p}}= \frac{4.8\times 10^8}{B^2 E^{\prime}_{\rm p,19}}  \,\, {\rm s}.
\end{equation}
Comparing this time with the escape time $t_{\rm esc}\sim R/\beta_{\rm esc}c$ we have that cooling in unimportant as far as $E_{\rm p,19}\lesssim1.4\times 10^4 \, \Gamma_1 B^{-2} R_{16}^{-1}\beta_{\rm esc}$, i.e. for energies well above those expected in the jet (assuming $\beta_{\rm esc}\sim 1$). 

The synchrotron luminosity from UHE protons can be expressed as (see Appendix):
\begin{equation}
\mathcal{L}_{\rm s,p}\simeq10^{41} B \left( \frac{\delta}{\Gamma} \right)^4 
\frac{P_{\rm CR,j}} {10^{43}\, {\rm erg \, s^{-1}}} \, E_{\rm max,19}^2 \,\,\, {\rm erg\,  s}^{-1}.
\label{lsprotsimple}
\end{equation}
For the values of the magnetic field considered here $B<10$ G this emission is therefore less important than or at least comparable to the SSC luminosity.






\section{SED modeling}

Besides the analytic estimates above, approximated but suitable to give an immediate glance of the parameter space, we also applied the one zone leptonic model fully described in Maraschi \& Tavecchio (2003) to reproduce the SED of the four EHBL. Since the shape and the position of the SSC peak is basically unconstrained (the only constraint being the upper limit to the luminosity), these models are only intended as an example and a consistency check of the previous treatment. Note however that, since the synchrotron peak is rather well described by the available data, the SED modeling can provide some interesting informations (in particular the shape of the electron distribution).

The emission region is assumed to be spherical with radius $R=10^{16}$ cm and filled with tangled magnetic field with intensity $B$. Emitting relativistic electrons are described by a smoothed broken power law energy distribution in the interval $\gamma_{\rm min}<\gamma<\gamma_{\rm max}$, with slopes $n_1$ and $n_2$ below and above the break at $\gamma_b$ and normalization $K$. Relativistic beaming effects are taken into account by the Doppler relativistic parameter, $\delta$.

For each source we calculate two models (see Fig.\ref{sed}), corresponding to $\delta=10$ and 30, having care to keep the SSC peak a factor of $\approx 10$ below the observed GeV-TeV data. Input parameters are reported in Table 1. 

Since in the framework of the hadronic cascade emission it is not required to reproduce the extremely hard de-absorbed TeV continuum (as done, e.g., in Tavecchio et al. 2011), we can relax the assumptions on the electron energy distribution. In particular we can avoid to assume an abrupt end of the electron energy distribution at a high $\gamma_{\rm min}\sim 10^4$. However, the UV-soft X-ray continuum is still constrained to be quite hard by the UVOT and XRT data, implying a rather flat distribution at low energy, with $p=1.5-1.7$. Such slopes are hardly compatible with the results of recent particle in cell simulations that indicates softer ($p>2$) electron spectra (e.g. Sironi \& Spitkovski 2011). However, as shown by Summerlin \& Baring (2012) through Monte Carlo simulations, considering oblique shocks, spectral indices down to $p\sim 1$ can be achieved. The magnetic field, very small in the previous models, has now values similar to or  larger than those inferred for the other, ``normal" BL Lac objects, $B\sim 0.1-1$ G. We  note that the break Lorentz factor we infer for the magnetic fields derived with $\delta=30$, $B\sim 0.2$ G, of the order of $\gamma _{\rm b}\simeq 10^4$, is close to what expected considering the cooling of electrons in the source light crossing time $R/c$, $\gamma_{\rm cool}\simeq 2\times 10^3 B^{-2}$. Therefore, only the electrons with Lorentz factors above $\gamma_{\rm b}$ are expected to cool, while at lower energies the electron distribution reflects the injection.

In these models the parameter $\gamma_{\rm min}$ is basically unconstrained. We fix $\gamma_{\rm min}=1$. With this choice the resulting electron and proton power are maximized. Similarly, $\gamma_{\rm max}$ is not well constrained. Future observations in the hard X-ray band (above 10 keV) with {\it Nustar} (Harrison et al. 2013) will surely allow us to better probe the synchrotron peak and infer the maximum energy of the accelerated electrons. 

In Table 2 we report the  powers carried by protons, electrons and magnetic field, the jet radiative power and the CR power, and the radiation and CR efficiencies (for negligible deflection). The last column reports $E_{\rm max}/Z$ calculated with the inferred physical quantities. The parameters are consistent with those derived analytically. Note the large efficiencies (both in radiation and UHECR), in some extreme cases close to 0.3-0.5,  required for $\delta=10$. Such large efficiency would imply a substantial jet (internal and/or kinetic) power dissipation.


\section{Discussion}

The synchrotron-SSC diagnostics robustly indicates that in principle EHBL can accelerate protons with energies larger than $2-3 \times 10^{19}$ eV. As shown by Murase et al. (2012), protons with such energies are only slightly deviated for the magnetic fields intensity and coherence lengths expected for cosmological filaments. The power required for the UHECR beams in this case can be easily supplied by the jet (Fig.\ref{emaxdelta}) with efficiencies $\eta_{\rm CR}\lesssim10\%$. 

In general, the efficiency required to power the UHECR beam is quite high for low Doppler factors ($\delta\sim10$). Since in this case the jet power is dominated by the magnetic field component (see Tab. 2), one could argue that the UHECR  energization involves some magnetic dissipation processes, such as magnetic reconnection (e.g., Giannios 2010). For larger bulk Lorentz factors the required CR efficiency rapidly decreases due to the narrower beaming cone.


We have assumed that the intrinsic SSC emission flux is $\approx$ 5-10 times lower than the hadronic cascade reprocessed flux, so that the contribution of the jet emission is almost negligible. The marginal evidence of mild (less than a factor of 2 in about one month) variability of the TeV flux of 1ES 0229+200, recently reported by  VERITAS (Cerruti et al. 2013), seems to indicate that the SSC intrinsic flux is not much smaller than what assumed here. In fact, in the framework of the hadronic cascade scenario, a quasi steady emission is expected: in the case of filaments, the time spread is expected to be of the order of $\Delta t\approx 10^3 E^{-2}_{19.5}$  yr (Murase et al. 2012, see also Prosekin et al. 2012), thus much longer than what observed . The only explanation for variability is therefore to assume a variable underlying SSC component, whose level is likely not much lower than what assumed in this work. Note that larger SSC luminosity would imply lower values of $E_{\rm max}$ and thus larger CR luminosities.

Finally, it is tempting to speculate about the possibility that a local (i.e. within the GZK radius, $\approx 100$ Mpc) population of misaligned EHBL contributes to the UHECR flux observed at the Earth (see Kotera \& Olinto 2011 for a review). 
Let us consider first a local population of EHBL characterized by {\it steady} emission. The density of sources required to account for the observed injected luminosity per unit volume, $\mathcal{U} \sim 10^{44}$ erg Mpc$^{-3}$ yr$^{-1}$ is $n_{\rm s}\simeq \mathcal{U}/2 P_{\rm CR,j}\sim 2 \mathcal{U} \Gamma^2 /L_{\rm CR}\sim 10^{-7} \, \Gamma^2_1$ Mpc$^{-3}$ (we used $L_{\rm CR}=5\times 10^{45}$ erg s$^{-1}$ and the factor 2 accounts for the existence of 2 jets). Even for $\Gamma=30$ the density is below the lower limit indicated by auto-correlation studies, $n_{\rm s}\gtrsim 10^{-5}$ Mpc$^{-3}$ (Kashti \& Waxman 2008).

Alternatively, one can consider sources characterized by activity periods of {\it finite duration} $t_{\rm dur}$.
In case of substantial deflection ($\theta _{\rm def}\approx 1$) of the injected UHECR  by a magnetized structure with spatial scale $D$, the time spread of UHECR is of the order of $\Delta T_{\rm def} \sim D/4c$ (e.g. Murase \& Takami 2009). If the activity duration is $t_{\rm dur}<< \Delta T_{\rm def}$, the flux would be diluted. Given sources of UHECR with intrinsic  luminosity $P_{\rm CR,j}$ and density $n_{\rm s}$, the measured luminosity per unit volume $\mathcal{U}$ would therefore be $\mathcal{U}\simeq P_{\rm CR,j} \, n_{\rm s} t_{\rm dur}/ \Delta T_{\rm def}$, from which the duration time of a single injection event would be  $t_{\rm dur}\sim 10^4 D_{1 {\rm Mpc}} n_{\rm s,-5}^{-1} \Gamma_1^2$  yr. This time is much smaller than the deflection time $D/c\sim 10^6 D_{1 {\rm Mpc}}$ yr, consistently with the assumption above. Importantly,  the required $t_{\rm dur}$ is also compatible with the minimum time of activity of EHBL requested in order to not increase the UHECR power in the hadronic cascade scenario. In fact, the expected times spread induced by the deviation in the EGMF (filaments) is $\Delta t_{\rm CR}\sim 2\times 10^4 E^{-2}_{19}$ yr, which is of the order of (or smaller than)  $t_{\rm dur}$ for $E=3\times 10^{19}$ eV,  $\Gamma \gtrsim10$ and $n_s$ compatible with the lower limits.

Since there are no EHBL within the GZK radius, local EHBL must be misaligned and thus the scenario above can be applied only if CR suffer significant deviation by the EGMF. At a first glance this seems to be in contradiction with the request that the UHECR beams of EHBL are not significantly deflected. A possibility to reconcile these two conditions is that, for some reasons, the environment of local EHBL is different than that of more distant sources. However, a more intriguing possibility is based on the observed change of composition of UHECR with energy, from protons to heavier ions (iron), occurring around $10^{19}$ eV (Abraham et al. 2010). Although this result, involving extrapolations of hadronic interaction parameters at energies unaccessible in laboratory (e.g. Kotera \& Olinto 2011), should be better understood, we note that if both protons and iron nuclei are accelerated by EHBL jets, a transition around $E\sim 2\times 10^{19}$ eV is expected in the scenario depicted above. In fact below these energies, both protons and iron nuclei are expected to be significantly deflected by the $\sim 10$ nG magnetic field of filaments. Above $E\sim 10^{19}$, however, the deflection for protons becomes negligible, while the heavier iron nuclei are still isotropized. The change in composition would therefore reflect the different deflection suffered by proton and iron in the EGMF magnetic field.

Concerning the misaligned counterparts, we note that EHBL are characterized by a very small radio luminosity, $L_{r}\lesssim10^{40}$ erg s$^{-1}$. Considering that this emission could be, at least partly, beamed, we can conclude that the local parent population belongs to the low-power tail of the weak (FR I) radiogalaxies, a (probably large) population of sources still not well studied (Baldi \& Capetti 2009).

In conclusion, our results are consistent -- in terms of maximum UHECR energy and sustainable luminosity -- with the assumptions of the hadronic cascade scenario for the $\gamma$-ray emission of EHBL if these sources do not reside in clusters and thus their UHE proton beams are not too much deflected by magnetic fields. The parameters are also consistent with the idea that a population of local EHBL can contribute to the UHECR flux measured at the Earth.

As argued by Murase et al. (2012) and Takami et al. (2013) a strong test of the model would be the detection of photons with energy above $\approx$ 20 TeV from EHBL. In fact, due to the absorption by EBL, an intrinsic $\gamma$-ray emission at these energies is expected to be completely suppressed, leaving as viable explanations the hadronic cascade model or the more speculative photon-axion like particle conversion scenario (e.g. De Angelis et al 2011). Future observations with CTA (Actis et al. 2011), or even with the planned ASTRI Mini-Array of small size telescopes (Di Pierro et al. 2013, Vercellone et al. 2013) can test and possibly confirm this prediction.

\section*{Acknowledgments}

I thank G. Morlino and G. Bonnoli for discussions and G. Ghisellini for a critical reading of the manuscript. I thank the referee for the constructive report. FT acknowledges financial contribution from grant PRIN-INAF- 2011. Part of this work is based on archival data provided by the ASI Science Data Center.

\appendix
\section{}

{\bf Derivation of the maximum energy in the Klein-Nishina limit}\\

In the KN limit, TMG98 derived the following relation between the observed quantities, the magnetic field $B$, the source radius $R$ and the Doppler factor:

\begin{equation}
B\delta^{1+\alpha_1}=\left( \frac{g}{\nu_C \nu _s} \right)^{\frac{1-\alpha_1}{2}} \left( \frac{2 L_s^2 f}{ R^2 \, c\, L_C}\right)^{\frac{1}{2}} \left( \frac{3\, m_{\rm e}c^2}{4 h}\right)^{1-\alpha_1}
\end{equation}
\noindent
where:
\begin{equation}
f=\frac{1}{1-\alpha_1}+\frac{1}{\alpha_2-1}
\end{equation}
\noindent
and:
\begin{equation}
g=\exp \left[ \frac{1}{1-\alpha_1}+\frac{1}{2(\alpha_2-\alpha_1)} \right].
\end{equation}

Rearranging the terms and using $E_{\rm max}=ZeBR\, \Gamma$ we can write:
\begin{equation}
E_{\rm max}=Z e\, \frac{\xi(\alpha_1,\alpha_2)}{\delta_1^{\alpha_1}} \, (\nu_{s} \nu_{C})^{\frac{\alpha_1-1}{2}}\,\frac{L_{s}}{L_{C}^{1/2}}\frac{\Gamma}{\delta} 
\end{equation}
\noindent
with  $\xi(\alpha_1,\alpha_2)$ given by:
\begin{equation}
\xi(\alpha_1,\alpha_2)=\left(\frac{2}{c} \right)^{\frac{1}{2}} f^{1/2} g^{(1-\alpha_1)/2} \left( \frac{3\, m_{\rm e}c^2}{4 h}\right)^{1-\alpha_1}.
\end{equation}
\noindent
\\

\noindent
{\bf Synchrotron emission from UHECR}\\

Let us now consider the synchrotron emission of the UHECR. The total luminosity is:
\begin{equation}
\mathcal{L}_{\rm s,p}\simeq\frac{4}{3} \sigma_{\rm T} c \left( \frac {m_{\rm e}}{m_{\rm p}} \right)^2 \frac{B^2}{8\pi} \mathcal{V} \delta^4 \int n_{\rm p}(\gamma_{\rm p})\gamma_{\rm p}^2 d\gamma_{\rm p}
\label{lsprot}
\end{equation}
\noindent
where the volume $ \mathcal{V}=(4/3)\pi R^3$.

Assuming a power law energy distribution $n_{\rm p}(\gamma_{\rm p})=k_{\rm p}\gamma_{\rm p}^{-p}$ with slope $p<3$ and limits $\gamma_{\rm p,1}<\gamma_{\rm p}<\gamma_{\rm p,2}$, the observed spectrum will peak at the frequency:
\begin{equation}
\nu _{s,p}\simeq 1.5 \times 10^{23} B E^{\prime \, 2}_{2,19} \,\, {\rm Hz},
\end{equation}
\noindent
where $E_2^{\prime}=\gamma_2 m_{\rm p}c^2$.

For definitiveness in the following we consider $p=2$

The proto-synchrotron luminosity can be estimated starting from the required UHECR (beaming-corrected) power, $P_{\rm CR,j}$ that can be expressed, similarly to the other forms of power, as: $P_{\rm CR,j}\simeq \pi R^2 U^{\prime}_{\rm CR} c \Gamma^2$, where, assuming a power law CR energy distribution with slope $p=2$:
\begin{equation}
U_{\rm CR}^{\prime}=m_{\rm p} c^2 \int n_{\rm p}(\gamma_{\rm p}) \, \gamma_{\rm p} \, d\gamma_{\rm p} = m_{\rm p} c^2 k_{\rm p} \ln \frac{E_{\rm p,2}}{E_{\rm p,1}}.
\label{uprot}
\end{equation}
Inserting this expression into $P_{\rm CR,j}$ we can derive $k_{\rm p}$ that, in turn, can be inserted into Eq. \ref{lsprot}. We obtain:
\begin{equation}
\mathcal{L}_{\rm s,p}\simeq \frac{2}{9\pi} \frac{\sigma_{\rm T}}{m_{\rm p}c^2} \left( \frac{m_{\rm e}}{m_{\rm p}}\right)^2 B^2 R \delta^2 \left( \frac{\delta}{\Gamma}\right)^2 \gamma_2 \frac{P_{\rm CR,j}}{\Lambda}
\end{equation}
\noindent
where $\Lambda = \ln (E_{\rm p,2}/E_{\rm p,1})$.
The radius can be eliminated using $E_{\rm max}=e B R \Gamma$. Further assuming $E_2=E_{\rm max}$ and inserting the numerical values we finally obtain:
\begin{equation}
\mathcal{L}_{\rm s,p}\simeq  10^{41} B \left( \frac{\delta}{\Gamma} \right)^4
\frac{P_{\rm CRj}}{10^{43} \, {\rm erg \, s^{-1}}} \frac{E_{\rm max,19}^2}{(\Lambda/20)} \,\,\,  {\rm erg \, s^{-1}}
\label{lsprotsimple}
\end{equation}
where $\Lambda\approx 20$.

For heavier ions, with charge $Ze$ and mass $m_{\rm i}=A m_{\rm p}$ and energy $E_{\rm max,i}=Z E_{\rm max}$ the synchrotron luminosity is
\begin{equation}
\mathcal{L}_{\rm s,i}= \frac{Z^6}{A^4} \mathcal{L}_{\rm s,p} \sim 30 \left(\frac{Z}{26}\right)^6 \left(\frac{A}{56}\right)^{-4} \mathcal{L}_{\rm s,p},
\end{equation}
\noindent
where the numerical values are specific for Fe nuclei. The peak frequency for iron is $\nu _{\rm s,i}= (Z/A)^3 \nu_{\rm s,p}\simeq 0.1 \nu_{\rm s,p}$.

\end{document}